\documentclass[a4paper]{jpconf}
\usepackage{graphicx}
\usepackage{amssymb,latexsym}

\def\be{\begin{equation}}
\def\ee{\end{equation}}
\def\bq{\begin{eqnarray}}
\def\eq{\end{eqnarray}}
\def\beq{\begin{eqnarray*}}
\def\eeq{\end{eqnarray*}}

\begin{document}
\title{Singularities of varying light speed cosmologies}

\author{John Miritzis$^{1,\dagger}$ and Spiros Cotsakis$^{2,\ddagger}$}
\address{$^1$Department of Marine Sciences, University of the Aegean,
University Hill, Mytilene 81100, Greece}
\address{$^2$Research Group of Mathematical Physics and Cosmology,
Department of Information and Communication Systems Engineering,
University of the Aegean, Karlovassi 83 200, Samos, Greece}
\ead{imyr@aegean.gr$\dagger$} \ead{skot@aegean.gr$\ddagger$}
\begin{abstract}
\noindent We study the possible singularities of isotropic
cosmological models that have a varying speed of light as well as
a varying gravitational constant. The field equations typically
reduce to two dimensional systems which are then analyzed both by
dynamical systems techniques in phase space and by applying the
method of asymptotic splittings. In the general case we find
initially expanding closed models which recollapse to a future
singularity and  open universes that are eternally expanding
towards the future. The precise nature of the singularities is
also discussed.
\end{abstract}

\section{Introduction}
Varying speed of light (VSL) cosmologies have recently received
considerable attention as alternatives to cosmological inflation
to provide a different basis for resolving the problems of the
standard model (see for example \cite{magu} and refs. therein).
Instead of adopting the inflationary idea that the very early
universe experienced a period of superluminar expansion, in these
universes one assumes that light travelled faster in the early
universe. In such cosmological models  the known puzzles of the
standard early universe cosmology are absent at the cost of
breaking the general covariance of the underlined gravity theory.
We do not enter here into a discussion on the foundations of  VSL
theories and the conceptual problems arising from the very meaning
of varying the speed of light (see \cite{magu} for a discussion).

In this paper we report on some preliminary results of our
on-going work \cite{cm07} on the application of the method of
asymptotic splittings introduced in \cite{cb06} to study the
singularities that may arise in VSL cosmologies. In particular, we
focus here on the VSL model proposed in \cite{alma} and further
investigated in \cite{barr}. An important characteristic of this
model is that one assumes minimal coupling at the level of the
Einstein equations so that a time-variable $c$ should not
introduce changes in the curvature terms in Einstein's equations
in the cosmological frame, Einstein's equations must still hold.
As Barrow points out \cite{barr}, $c$ changes in the local
Lorentzian frames associated with the cosmological expansion, a
special-relativistic effect, so that the resulting theory is not
covariant and one has to make a specific choice of time
coordinate. Choosing that specific time to be comoving proper time
the Friedman equations retain their form with $c(t)$ and $G(t)$
varying.

We assume an equation of state of the form $p=(\gamma-1)\rho c^{2},$ with $%
0<\gamma \leq2$ and we write the Friedman equations with varying
$c\left( t\right) $ and $G\left( t\right) $ as follows:
\begin{equation}
\left( \frac{\dot{a}}{a}\right) ^{2}+\frac{kc^{2}}{a^{2}}=\frac{8\pi G}{3}%
\rho,  \label{frie}
\end{equation}%
\begin{equation}
\frac{\ddot{a}}{a}=-\frac{8\pi G}{6}\left( 3\gamma-2\right) \rho .
\label{rayc}
\end{equation}
Here  $a$ is the scale factor, $k=0,+1,$ or $-1.$ Differentiating
the first and using the second equation above we obtain the
conservation equation
\begin{equation}
\dot{\rho}+3\gamma \rho \frac{\dot{a}}{a}=-\rho \frac{\dot{G}}{G}+3\frac{k}{%
a^{2}}\frac{c\dot{c}}{4\pi G},  \label{cons}
\end{equation}
and setting  $a=x,$ $\dot{x}=y$ we obtain the non-autonomous system%
\begin{eqnarray}
\dot{x} & =&y,  \nonumber \\
\dot{y} & =&-\frac{8\pi G\left( t\right) }{6}\left( 3\gamma-2\right) x\rho,
\label{nona} \\
\dot{\rho} & =&-3\gamma \frac{y}{x}\rho-\rho \frac{\dot{G}\left( t\right) }{%
G\left( t\right) }+3\frac{k}{x^{2}}\frac{c\left( t\right) \dot{c}\left(
t\right) }{4\pi G\left( t\right) },  \nonumber
\end{eqnarray}
subject to the constraint
\[
\frac{y^{2}}{x^{2}}+\frac{kc^{2}\left( t\right) }{x^{2}}=\frac{8\pi G\left(
t\right) }{3}\rho.
\]

\section{The flat case}
For flat ($k=0$) models we set $z=G\rho$ and the system
(\ref{nona}) becomes
\begin{eqnarray}
\dot{x} & =&y,  \nonumber \\
\dot{y} & =&-\frac{8\pi}{6}\left( 3\gamma-2\right) xz, \nonumber\\
\dot{z}+3\gamma\frac{y}{x}z&=&0,\label{flat}\end{eqnarray}
subject to the constraint%
\[
\frac{y^{2}}{x^{2}}=\frac{8\pi}{3}z.
\]
This system is exactly that which was analyzed using the method of
asymptotic splittings in \cite{cb06},  and we rewrite their
result. Putting
\[
x=\alpha \tau^{p},\ y=\beta \tau^{q},\ z=\delta \tau^{r},
\]
we find the balance
\begin{equation}
p=\frac{2}{3\gamma},\ q=p-1,\ r=-2,\ \
\beta=\frac{2}{3\gamma}\alpha ,\delta=\frac{1}{6\pi \gamma^{2}},\
\alpha=\mathrm{arbitrary.}  \label{bala}
\end{equation}
Computation of the eigenvalues of the Kovalevskaya matrix
$K=D\mathbf{f} \left( \mathbf{\alpha}\right) -diag\left(
\mathbf{p}\right) $ yields the values $-1,0,\frac{2\left(
3\gamma-2\right) }{3\gamma}.$ Applying the constraint to the
second equation we obtain an integrable system with general
solution
\begin{equation}
x  =(At+C)^{2/3\gamma},\quad y
={\frac{2A}{3\gamma}}(At+C)^{2/3\gamma-1}, \quad    z
={\frac{A^{2}}{6\pi \gamma^{2}}}(At+C)^{-2},\label{exact}
\end{equation}
where $A$ and $C$ are constants of integration. We see that the
density function always has a finite time singularity. On the
other hand, the Hubble parameter blows up in finite time only for
$\gamma<2/3.$ It is interesting that the leading-order terms
(\ref{bala}) obtained from the calculation of the dominant balance
already contain all the information provided by the exact solution
(\ref{exact}), an effect due to the fact that the system is
weight-homogeneous. We will report on the details of the full
analysis of this system elsewhere.

\section{Reduction to two dimensions}

We assume a power-law dependence of the function $c,$
\[
c=c_{0}a^{n},\ \ \ n\in \mathbb{R},\ \ G=\mathrm{constant},
\]%
and  use a system of units with $8\pi G=1=c_{0}^{2}.$ We can avoid
having to deal with denominators if we
set $x=1/a,$ $H=\dot{a}/a.$ Then the system (\ref{frie}), (\ref{rayc}), (\ref%
{cons}) becomes
\begin{eqnarray}
\dot{x} &=&-xH,  \nonumber \\
\dot{H} &=&-H^{2}-\frac{\left( 3\gamma -2\right) }{6}\rho ,  \label{3-d} \\
\dot{\rho} &=&-3\gamma \rho H+6knu^{2-2n}H,  \nonumber
\end{eqnarray}%
and  the constraint reads
\begin{equation}
H^{2}+ku^{2-2n}=\frac{1}{3}\rho .  \label{const}
\end{equation}%
We use the constraint to eliminate $x,$ and write the system in
the form
\begin{eqnarray}
\dot{H} &=&-H^{2}-\frac{\left( 3\gamma -2\right) }{6}\rho ,  \nonumber \\
\dot{\rho} &=&-\left( 3\gamma -2n\right) \rho H-6nH^{3}.  \label{2-dim}
\end{eqnarray}%
From Eq.  (\ref{const}) it follows that the phase space of
(\ref{2-dim}) is the set $ \left\{ \left( H,\rho \right) \in
\mathbb{R}^{2}:\rho \geq 0,\rho -3H^{2}>0\right\} $ for closed
models, and $\left\{ \left( H,\rho \right) \in \mathbb{R}^{2}:\rho
\geq 0,\rho -3H^{2}<0\right\} $ for the open ones. Note that the
corresponding  equations in general relativity have the feature
that the conservation equation corresponding to the second equation in  (\ref%
{2-dim}) is just $\dot{\rho}=-3\gamma \rho H,$ which implies that the line $%
\rho=0 $ is invariant, i.e., the trajectories of the system cannot
cross the line $\rho=0.$ On the other hand, here equations
(\ref{2-dim}) without further assumptions  do not guarantee that a
solution curve starting at a point with $\rho>0$ will not
eventually enter the region with $\rho<0,$ which of course is
unphysical.

A first task of the method of asymptotic splittings is to find all
possible asymptotic forms $ H=\alpha \tau ^{p},\, \rho =\beta \tau
^{q} $ admitted by the system (\ref{2-dim}).  One balance gives
 $ p=-1,q=-\left( 3\gamma -2n\right) $ with  $\alpha =1,\,\beta =\mathrm{arbitrary}.$
This has K-exponents $(-1,0)$. A second interesting balance is for
$p=-1,q=-2$ with coefficients given by
\[
\left( \alpha =\frac{2}{3\gamma },\beta =\frac{4}{3\gamma ^{2}}\right)
,\left( \alpha =\frac{1}{1-n},\beta =-\frac{6n}{\left( 3\gamma -2\right)
\left( 1-n\right) ^{2}}\right) .
\]%
The K-exponents are in this case given by the forms
\[
\left( -1,\frac{2}{3\gamma}\left( 3\gamma+2n-2\right) \right) ,\left( -1,%
\frac{3\gamma+2n-2}{n-1}\right) .
\]
These results lead to formal series expansions of the solutions in
a suitable neighborhood of the finite time singularity, they will
be presented elsewhere, \cite{cm07}. It is interesting that a
phase space analysis shows that only closed models run into a
finite time future singularity. In fact the phase portrait of
(\ref{2-dim}) with $n=-1/2,\gamma =1$ is shown in Figure 1.
\begin{figure}[htb]
\begin{center}
\includegraphics{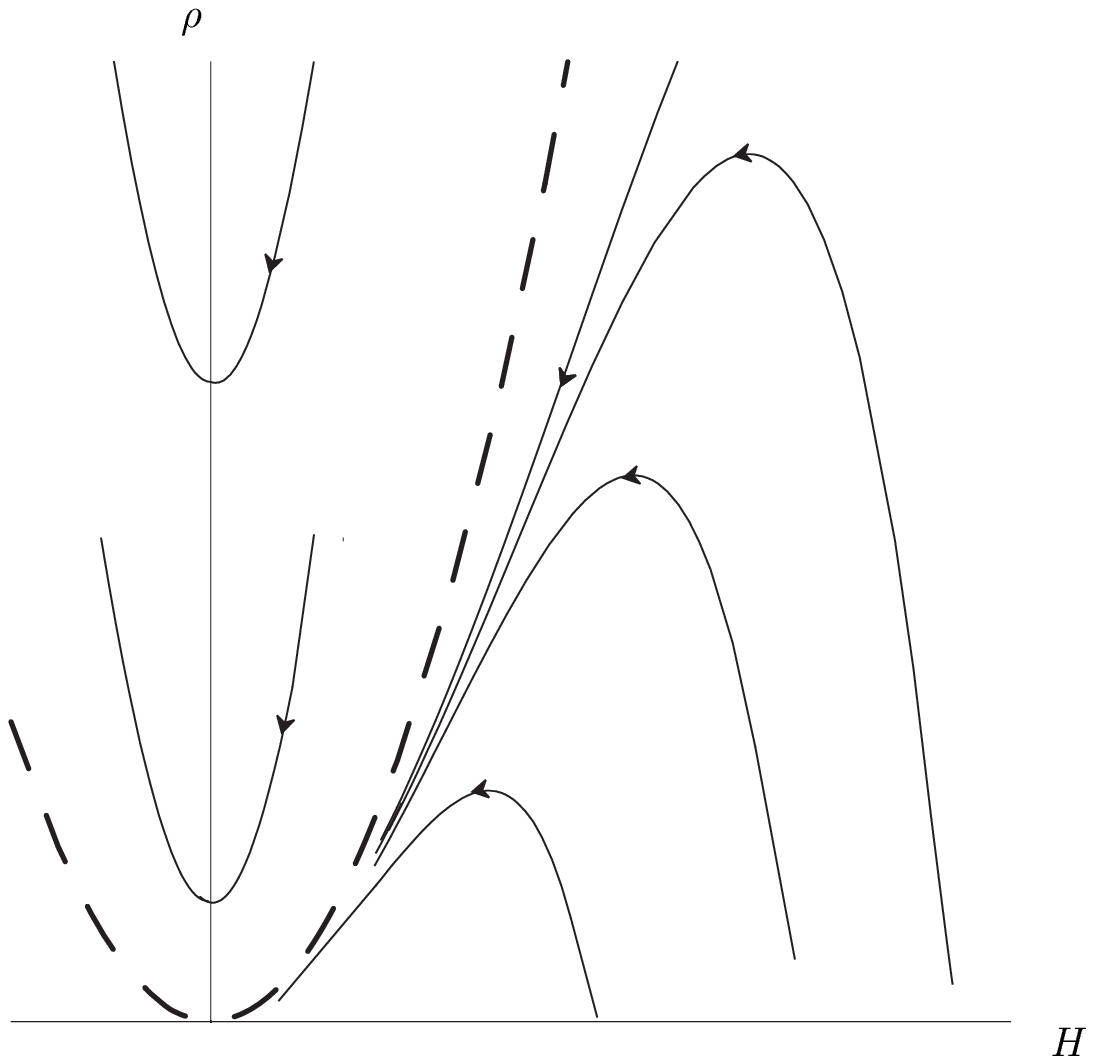}
\end{center}
\caption{The phase portrait of (\protect\ref{2-dim})}
\label{fig1}
\end{figure}
The dashed line in this Figure separates the closed from the open
models. We see that every initially expanding universe starting
below this line eventually approaches the origin which corresponds
to zero density and expansion rate. On the other hand, initially
expanding universes starting above the dashed line eventually
contract and $H\rightarrow -\infty ,$ $\rho \rightarrow \infty $
in a finite time. Numerical experiments confirm the above
analysis.

\section{The general case}
The reduction of the three dimensional dynamical system
(\ref{3-d}) to the two-dimensional system (\ref{2-dim}) is not
unique. We can use the constraint (\ref{const}) to eliminate $\rho
$ instead of $u.$ However, we shall see that assuming a power-law
dependence for $G$ as well, we can handle both cases at once. In
this Section we assume that
\[
c=c_{0}a^{n},\ \ \ G=G_{0}a^{m},\ \ \ n,m\in \mathbb{R},
\]%
and use a system of units with $8\pi G_{0}=1=c_{0}^{2}.$ Setting
again $x=1/a,$ $H=\dot{a}/a,$ the system (\ref{frie}), (\ref{rayc}), (\ref%
{cons}) can be written as
\begin{eqnarray}
\dot{x} &=&-xH, \\
\dot{H} &=&-H^{2}-\frac{1}{6}\left( 3\gamma -2\right) \rho x^{-m},  \nonumber
\\
\dot{\rho} &=&-3\gamma \rho H-m\rho H+6knx^{2+m-2n}H,  \nonumber
\end{eqnarray}%
subject to the constraint%
\begin{equation}
H^{2}+kx^{2-2n}=\frac{1}{3}\rho x^{-m}.  \label{constr}
\end{equation}%
We use the constraint to eliminate $\rho ,$ and arrive at the
two-dimensional system
\begin{eqnarray}
\dot{x} &=&-xH,  \nonumber \\
\dot{H} &=&-\frac{3\gamma }{2}H^{2}-\frac{\left( 3\gamma -2\right) }{2}%
kx^{2-2n}.  \label{2-di}
\end{eqnarray}%
Note that from (\ref{constr}) the phase space of (\ref{2-di}) is the set
\begin{equation}
\left\{ \left( x,H\right) \in \mathbb{R}^{2}:x\geq 0,\
H^{2}+kx^{2-2n}>0.\right\} \newline
\label{phsp}
\end{equation}
This system is now in a form suitable for both a dynamical systems
analysis and an application of the method of asymptotic
splittings. Full details will be given elsewhere but we report
here on some partial results to give a flavor of the analysis.
Putting $x=\alpha \tau ^{p},\, H=\beta \tau ^{q}$ we find a
possible balance with $q=-1,p=-2/3\gamma  $ and $\beta =2/3\gamma
=1,\alpha =\mathrm{arbitrary},$ acceptable for $n<0$. The
Kovalevskaya eigenvalues for this balance are at $-1,0,$
compatible with the arbitrariness of $\alpha .$

In the case where all terms are dominant,  the vector field is
\[
\mathbf{f}^{(0)}=\left[
\begin{array}{c}
-xH \\
-\frac{3\gamma }{2}H^{2}-\left( \frac{3\gamma }{2}-1\right)
kx^{2-2n}
\end{array}
\right]
\]%
and we find the balance (disregarding the case $\alpha =0,\beta
=0,$)
\[
\left( q=-1,\ p=\frac{1}{n-1}\right) ,\ \ \left( \beta
=\frac{1}{1-n},\ \ \alpha ^{2-2n}=\frac{2-2n-3\gamma }{\left(
3\gamma -2\right) k\left( 1-n\right) ^{2}}\right),
\]
with K-exponents given by
\[
-1,\frac{2-2n-3\gamma}{1-n}.
\]
Note that  if $n=-1/2$ and $\gamma=1$ (dust), or if $n=-1$ and
$\gamma=4/3$ (radiation) the second eigenvalue is equal to $0$.

We may proceed to analyze the system (\ref{2-di}) in phase space.
First, we write the system it as a single differential equation,
\begin{equation}
\frac{dH}{dx}=\frac{3\gamma H^{2}+\left( 3\gamma -2\right) kx^{2-2n}}{2xH},
\label{de}
\end{equation}
and we set $H^{2}=z$ to obtain a linear differential equation for
$z$ which is easily integrable. We find that (\ref{de}) has the
general solution
\begin{eqnarray}
H^{2} &=&\frac{\left( 3\gamma -2\right) k}{2-3\gamma -2n}x^{2-2n}+Cx^{3%
\gamma }\ \ \ \textrm{if\ \ }3\gamma +2n\neq 2,  \label{gens} \\
H^{2} &=&\left( 3\gamma -2\right) k\,x^{3\gamma }\ln x+Cx^{3\gamma
}\ \ \ \textrm{if\ \ }3\gamma +2n=2,  \nonumber
\end{eqnarray}%
where $C$ is a constant of integration. Suppose that $3\gamma +2n\neq 2.$ We
observe that the leading-order terms of the first decomposition, namely
\[
x=\alpha \tau ^{-2/3\gamma },\ \ \ H^{2}=\frac{4}{9\gamma
^{2}}\tau ^{-2},
\]%
reproduce the $Cx^{3\gamma }$ part of the solution (\ref{gens}). The
leading-order behavior of the last decomposition,
\[
x\sim \tau ^{1/\left( n-1\right) },\ \ \ H^{2}\sim \tau ^{-2},
\]%
recovers the $\frac{\left( 3\gamma -2\right) k}{2-3\gamma -2n}x^{2-2n}$ part
of the solution.

\begin{figure}[htb]
\begin{center}
\includegraphics{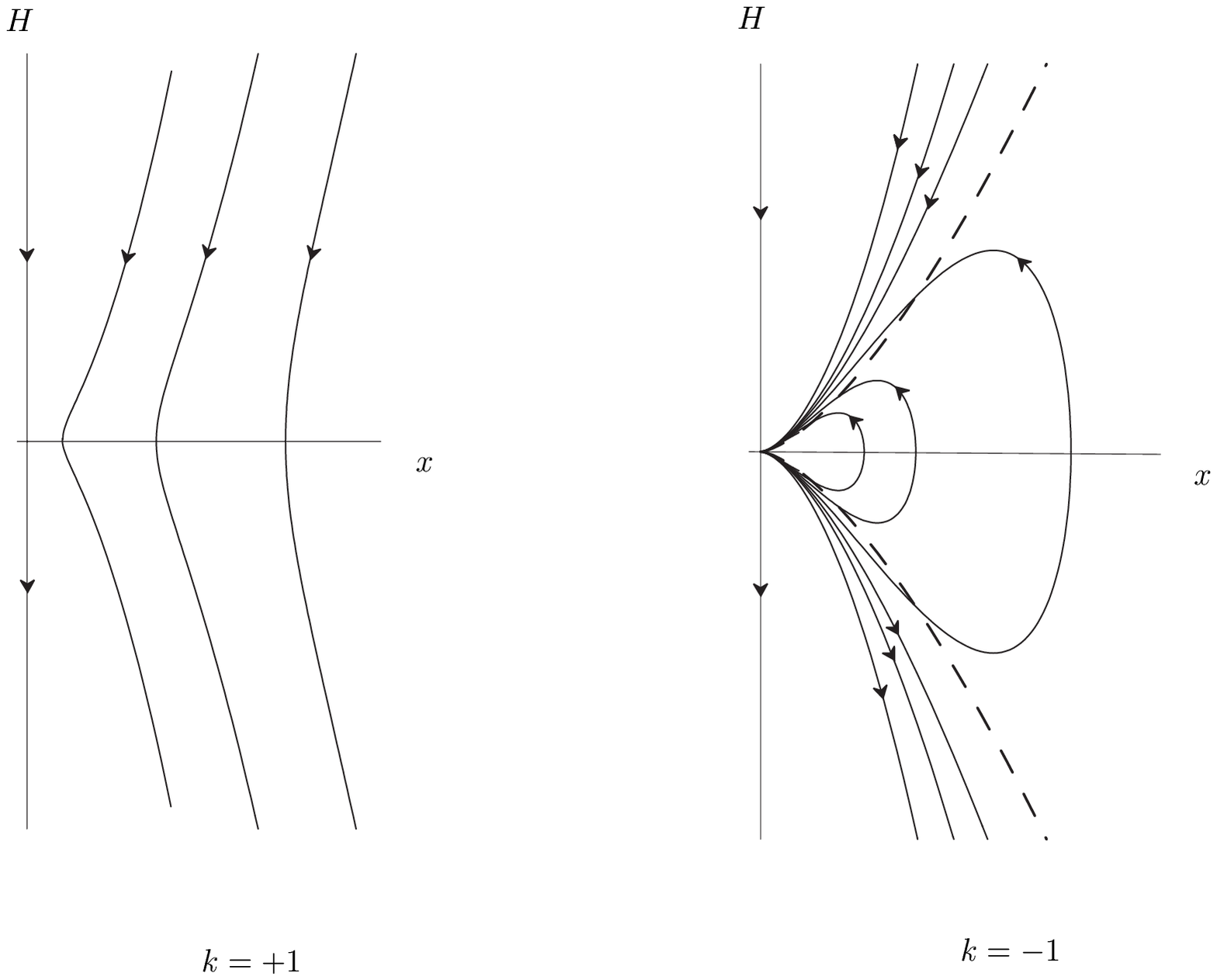}
\end{center}
\caption{The phase portrait of (\protect\ref{2-di})}
\label{fig2}
\end{figure}

The integral curves of (\ref{gens}) allow us to sketch the phase
portrait of (\ref{2-di}). Let us assume for concreteness that
$n=-1,$ $\gamma=4/3$ (radiation). There are two cases to consider:

\begin{itemize}
\item Closed models. The phase space (see (\ref{phsp})) is the half-plane $%
x\geq 0.$ Equation (\ref{gens}) yields $H^{2}=-2x^{3}+Cx^{4},$
which implies that $C>0.$ Any orbit starting in the first quadrant
satisfies $x\geq 2/C,$ i.e., there are no solutions approaching
the origin. For $C>0,$ any orbit of (\ref{2-di}) starting in the
first quadrant crosses the $x-$axis at $2/C$ and remains in the
fourth quadrant. We conclude that any initially expanding closed
universe reaches  maximum expansion at some finite time and then
recollapse begins, i.e. $H$ approaches $-\infty $ in a finite
time. The phase portrait is shown in Figure 2.

\item Open models. From (\ref{phsp}) we see that the phase space is the set $%
x\geq 0,\ H^{2}>x^{3}$. In the first quadrant this set is represented as the
area above the dashed line in Figure 2 B. Eq. (\ref{gens}) yields $%
H^{2}=+2x^{3}+Cx^{4},$ which implies that for $C>0$ any orbit
starting in the first quadrant asymptotically approaches the
origin. For $C<0,$ there are homoclinic curves connecting the
origin with itself. However, we must remember that the allowed
initial conditions for expanding universes lie above the dashed
line in Figure 2. It can be shown that these trajectories also
asymptotically approach $\left( 0,0\right) .$ We conclude that any
initially expanding open universe remains ever-expanding.
\end{itemize}

In all, we find that VSL cosmological models may share many
interesting dynamical characteristics not fully present in the
more conventional models so that a more detailed study of the
singularities in these universes is worthwhile as a future
challenge.

\end{document}